\documentstyle[12pt]{article}

\def\hybrid{\topmargin -20pt	\oddsidemargin 0pt
	\headheight 0pt	\headsep 0pt
	\textwidth 6.25in	
	\textheight 9.5in	
	\marginparwidth .875in
	\parskip 5pt plus 1pt	\jot = 1.5ex}

\hybrid
\def\baselinestretch{1.2}
\catcode`\@=11
\def\marginnote#1{}
\newcount\hour
\newcount\minute
\newtoks\amorpm
\hour=\time\divide\hour by60
\minute=\time{\multiply\hour by60 \global\advance\minute by-\hour}
\edef\standardtime{{\ifnum\hour<12 \global\amorpm={am}%
	\else\global\amorpm={pm}\advance\hour by-12 \fi
	\ifnum\hour=0 \hour=12 \fi
	\number\hour:\ifnum\minute<10 0\fi\number\minute\the\amorpm}}
\edef\militarytime{\number\hour:\ifnum\minute<10 0\fi\number\minute}
\def\draftlabel#1{{\@bsphack\if@filesw {\let\thepage\relax
   \xdef\@gtempa{\write\@auxout{\string
      \newlabel{#1}{{\@currentlabel}{\thepage}}}}}\@gtempa
   \if@nobreak \ifvmode\nobreak\fi\fi\fi\@esphack}
	\gdef\@eqnlabel{#1}}
\def\@eqnlabel{}
\def\@vacuum{}
\def\draftmarginnote#1{\marginpar{\raggedright\scriptsize\tt#1}}

\def\draft{\oddsidemargin -.5truein
	\def\@oddfoot{\sl preliminary draft \hfil
	\rm\thepage\hfil\sl\today\quad\militarytime}
	\let\@evenfoot\@oddfoot	\overfullrule 3pt
	\let\label=\draftlabel
	\let\marginnote=\draftmarginnote
   \def\@eqnnum{(\theequation)\rlap{\kern\marginparsep\tt\@eqnlabel}%
\global\let\@eqnlabel\@vacuum}  }

\def\preprint{\twocolumn\sloppy\flushbottom\parindent 2em
	\leftmargini 2em\leftmarginv .5em\leftmarginvi .5em
	\oddsidemargin -.5in	\evensidemargin -.5in
	\columnsep .4in	\footheight 0pt
	\textwidth 10.in	\topmargin  -.4in
	\headheight 12pt \topskip .4in
	\textheight 6.9in \footskip 0pt
	\def\@oddhead{\thepage\hfil\addtocounter{page}{1}\thepage}
	\let\@evenhead\@oddhead	\def\@oddfoot{}	\def\@evenfoot{} }

\def\numberbysection{\@addtoreset{equation}{section}
	\def\theequation{\thesection.\arabic{equation}}}

\def\underline#1{\relax\ifmmode\@@underline#1\else
	$\@@underline{\hbox{#1}}$\relax\fi}

\def\titlepage{\@restonecolfalse\if@twocolumn\@restonecoltrue
\onecolumn
     \else \newpage \fi \thispagestyle{empty}\c@page\z@
	\def\thefootnote{\fnsymbol{footnote}} }

\def\endtitlepage{\if@restonecol\twocolumn \else \newpage \fi
	\def\thefootnote{\arabic{footnote}}
	\setcounter{footnote}{0}}  

\catcode`@=12
\relax

\def\figcap{\section*{Figure Captions\markboth
	{FIGURECAPTIONS}{FIGURECAPTIONS}}\list
	{Figure \arabic{enumi}:\hfill}{\settowidth\labelwidth{Figure

999:}
	\leftmargin\labelwidth
	\advance\leftmargin\labelsep\usecounter{enumi}}}
 \relax
\def\tablecap{\section*{Table Captions\markboth
	{TABLECAPTIONS}{TABLECAPTIONS}}\list
	{Table \arabic{enumi}:\hfill}{\settowidth\labelwidth{Table

999:}
	\leftmargin\labelwidth
	\advance\leftmargin\labelsep\usecounter{enumi}}}
 \relax
\def\reflist{\section*{References\markboth
	{REFLIST}{REFLIST}}\list
	{[\arabic{enumi}]\hfill}{\settowidth\labelwidth{[999]}
	\leftmargin\labelwidth
	\advance\leftmargin\labelsep\usecounter{enumi}}}
 \relax
\makeatletter
\newcounter{pubctr}
\def\publist{\@ifnextchar[{\@publist}{\@@publist}}
\def\@publist[#1]{\list
	{[\arabic{pubctr}]\hfill}{\settowidth\labelwidth{[999]}
	\leftmargin\labelwidth
	\advance\leftmargin\labelsep
	\@nmbrlisttrue\def\@listctr{pubctr}
	\setcounter{pubctr}{#1}\addtocounter{pubctr}{-1}}}
\def\@@publist{\list
	{[\arabic{pubctr}]\hfill}{\settowidth\labelwidth{[999]}
	\leftmargin\labelwidth
	\advance\leftmargin\labelsep
	\@nmbrlisttrue\def\@listctr{pubctr}}}
 \relax
\makeatother
\newskip\humongous \humongous=0pt plus 1000pt minus 1000pt
\def\caja{\mathsurround=0pt}
\def\eqalign#1{\,\vcenter{\openup1\jot \caja
	\ialign{\strut \hfil$\displaystyle{##}$&$
	\displaystyle{{}##}$\hfil\crcr#1\crcr}}\,}
\newif\ifdtup

\relax

\def\thefootnote{\fnsymbol{footnote}}
\def\be{\begin{equation}}
\def\ee{\end{equation}}
\def\ba{\begin{eqnarray}}
\def\ea{\end{eqnarray}}
\def\d{\partial}

\def\s{\sigma}

\def\P{\Phi}

\def\ub{{\bar u}}
\begin{document}
\renewcommand{\theequation}{\thesection.\arabic{equation}}
\begin{titlepage}
\begin{center}

\hfill CERN-TH.7121/93\\
\hfill HUB-IEP-93/8\\
\hfill LPTENS 93/51\\
\hfill hep-th/9312143\\
\vskip .1in

{\large \bf Non-compact Calabi-Yau Spaces
and other Non-Trivial Backgrounds for
Four-dimensional Superstrings}
\vskip .5in

{\bf E. Kiritsis, C. Kounnas\footnote{On leave from Ecole Normale
Sup\'erieure,
24, rue Lhomond, 75231 Paris Cedex 05, France.} }
\vskip .1in
{\em CERN, Geneva, SWITZERLAND}
\vskip .1in
and
\vskip .1in
{\bf D. L\"ust}

\vskip .1in
{\em Humboldt Universit\"at zu Berlin\\
Fachbereich Physik\\
D-10099 Berlin, GERMANY}

\end{center}

\vskip .7in

\begin{center} {\bf ABSTRACT } \end{center}
\begin{quotation}\noindent
A large class of new 4-D superstring vacua
with non-trivial/singular geometries, spacetime supersymmetry
and other background fields (axion, dilaton) are found.
Killing symmetries are generic and are associated with
non-trivial dilaton and
antisymmetric
tensor fields. Duality symmetries preserving N=2 superconformal
invariance
are employed to generate a large class of explicit metrics for
non-compact
4-D Calabi-Yau manifolds with Killing symmetries.
\end{quotation}
\vskip 1cm

\begin{center}
{\em To appear in ``Essays on Mirror Manifolds II''}
\end{center}

\vfill
\noindent
CERN-TH.7121/93\\
December 1993\\
\end{titlepage}
\vfill
\eject
\def\baselinestretch{1.2}
\baselineskip 16 pt
\noindent
\section{Introduction}
\setcounter{equation}{0}

Ricci-flat K\"ahler spaces \cite{kahler}, so-called Calabi-Yau
spaces,
provide consistent backgrounds \cite{4d} for the propagation of
superstrings or heterotic strings. These backgrounds
lead to target space suprsymmetry and, consequently, the perturbative
vacuum is
guaranteed to be stable. Moreover,
during recent years many of the Calabi-Yau backgrounds were shown to
correspond to exact $N=2$ superconformal field theories \cite{Ge}.
In the past the discussion
mainly concentrated on flat four-dimensional Minkowski space-time
tensored with a six-dimensional internal compact K\"ahler space
without torsion
and with constant dilaton field.
We  would like however, to construct supersymmetric string vacua
with, in addition
to the metric background, more general (non-constant) background
fields. Moreover,
to address certain important questions in quantum gravity one has to
consider string backgrounds which describe four-dimensional curved
and non-compact
space-times. In particular, one is interested to construct exact
superconformal field theories which correspond to
four-dimen\-sio\-nal
black-hole backgounds,  cosmological or
supersymmetric instanton type of solutions.

In this contribution  we will report about a relatively systematic
discussion \cite{kkl}
on
supersymmetric string backgrounds with $N=2$ or $N=4$ superconformal
symmetry,
based on compact as well as
non-compact spaces plus
non-trivial antisymmetric tensor-field
and non-con\-stant dilaton.
Thus, we will extend in a more systematic way
the exact N=4 solution constructed recently, \cite{Ca,C2}.
In contrast to the compact Calabi-Yau spaces, almost all backgrounds
with non-trivial dilaton field will possess Killing symmetries.
Many such backgrounds exhibit singularities on some hypersurface
in spacetime and
 can be regarded as a higher dimensional
 generalization of the two-dimensional black-hole considered
in \cite{bh}.

A key to the proper understanding of string propagation
on curved spaces is provided by duality symmetries present in curved
backgrounds \cite{B,K,R,GK,super}.
Duality symmetries relate  different backgrounds which
nevertheless
correspond to the same (super)conformal field theory.
The duality
symmetries become manifest after a rearrangement in the Hilbert-space
of the superconformal field theory and mean that stringy probes,
when excited in different modes, see different geometries or
topologies. Therefore, the concept of geometry or topology
is not well defined in string theory.
For the case of compact (Calabi-Yau) backgrounds, several very
intruiging
examples of stringy duality equivalences were found. The simplest
example
of duality symmetries is the $R\rightarrow 1/R$ transformation for
toroidal type
of backgrounds \cite{flat} where $R$ is the characteristic length
scale of the compact space.
Here the duality symmetry originates from the exchange of  internal
momentum and winding
modes. A second, very interesting example where duality plays a
central role
is mirror symmetry
for a general class of Calabi-Yau compactifications \cite{mirror}.
Mirror symmetry
exchanges compact spaces of different topology but the same string
physics.
Duality symmetries also persist
in several non-compact backgrounds, where momentum states are
exchanged by
oscillator type of excitations \cite{K}.
We expect that for non-compact spaces, the existence of the duality
symmetries will radically modify our understanding of space-time at
least in regions
of very strong curvature.
In this contribution we will embark into finding four-dimensional
backgrounds for superstring propagation.
We will demand that the worldsheet theory has at least N=2
superconformal invariance in order to hope for spacetime
supersymmetry (due to the presence of spectral flow which will pair
the bosonic and fermionic spectrum).
We will then analyze the one-loop $\beta$-function equations and find
many interesting solutions (duality will be one of our tools).
Among other things we will see that some interesting
non-K\"ahlerian $ N=4$ solutions, which describe four-dimensional
axionic instantons, are dual-equivalent to
four-dimensional, non-compact
Ricci-flat K\"ahler spaces.

\renewcommand{\theequation}{\thesection.\arabic{equation}}

\section{The $N=2$ ($N=4$) Background and $U(1)$ Duality
Transformations}

The most general $N=2$ superspace action \cite{GHR}
for $m$ chiral superfields $U_i$ ($i=1,\dots , m$) and
$n$ twisted chiral superfields $V_p$ ($p=1,\dots ,n$)
in two dimensions is determined by a single real function
$K(U_i,\bar U_i,V_p,\bar V_p)$ (which we will henceforth call the
quasi-K\"ahler potential):
\be
S={1\over 2\pi \alpha '}\int{\rm d}^2xD_+D_-\bar D_+\bar D_-
K(U_i,\bar U_i,V_p,\bar V_p).\label{action}
\ee
The fields $U_i$ and $V_p$ obey a chiral or twisted chiral
constraint
\be
\bar D_\pm U_i=0,\qquad\bar D_+V_p=D_-V_p=0.\label{constraint1}
\ee
The action (\ref{action}) is invariant, up to total derivatives,
under quasi-K\"ahler gauge transformations
\be
K\rightarrow K+f(U_i,V_p)+g(U_i,\bar V_p)
+\bar f(\bar U_i,\bar V_p)+\bar g(\bar U_i,V_p).\label{inva}
\ee

To see the background interpretation of the theory it is convenient
to write down the purely bosonic part of the superspace action
(\ref{action}):
\be
\eqalign{
S=-{1\over 2\pi\alpha '
}\int{\rm d}^2x&\lbrack K_{u_i\bar u_j}\partial^a u_i
\partial_a\bar u_j-K_{v_p\bar v_q}\partial^a v_p\partial_a\bar v_{ q}
\cr& +\epsilon_{ab}(K_{u_i\bar v_p}\partial_a u_i\partial_b\bar v_{
p}
+K_{v_p\bar u_i}\partial_a v_p\partial_b\bar u_{i})\rbrack,
\label{bosonic}\cr}
\ee
where
$K_{u_i\bar u_j}={\partial^2K\over\partial U_i\partial\bar
U_j}$, etc.
Here $u_{i}$ is the lowest component of the superfield $U_{i}$ and so
on.
Thus, one recognizes that the first two terms in above equation
describe the  in general non-K\"ahlerian metric background of the
model
(the metric is K\"ahler only when $m=0$ or $n=0$). The
$\epsilon_{ab}$-term in (\ref{bosonic}) provides the
antisymmetric
tensor field background.

It follows that the field strength $H_{\mu\nu\lambda}
=\partial_\mu B_{\nu\lambda}
+\partial_\nu B_{\lambda\mu}
+\partial_\lambda B_{\mu\nu}$
can also be expressed entirely in terms
of the function $K$:
$H_{u_i\bar u_jv_p}
=
{\partial^3K\over\partial U_i\partial\bar U_j\partial V_p}$, etc.

Of course, in order that these backgrounds provide
consistent string solutions, they have to satisfy the string equation
of motion, i.e. the vanishing of the $\beta$-function equations
\cite{beta}.
Including also the dilaton background $\Phi(u_i,v_p)$,
we obtain  the following equations of motion for the background
fields,
\be
\eqalign{0&=\beta_{\mu\nu}^G=R_{\mu\nu}-{1\over
4}H_\mu^{\lambda\sigma}
H_{\nu\lambda\sigma}+2\nabla_\mu\nabla_\nu\Phi +O(\alpha')\cr
0&=\beta_{\mu\nu}^B=\nabla_\lambda
H_{\mu\nu}^\lambda-2(\nabla_\lambda
\Phi )H_{\mu\nu}^\lambda+O(\alpha').\label{betaf}\cr}
\ee
These equations will lead to some differential equations
for the two functions $K$ and $\Phi$ as we will discuss in the
following.
Moreover, the central charge deficit $\delta c$ of the
background  is determined by the $\beta$-function of the dilaton
field as
\be
\delta c\equiv c-{3D\over 2}={3\over 2}\alpha'
\lbrack 4(\nabla\Phi)^2-4\nabla^2\Phi-R+{1\over 12}
H^2\rbrack+O(\alpha'^{2}).\label{dilaton}
\ee
We must emphasize here that in the presence of N=4 superconformal
symmetry the
solution to the lowest order in $\alpha'$ is exact to all orders
in a specific scheme, and $\delta c$ remains zero to all orders
\cite{papa}.

We consider without loss of generality the simplest case of a single
$U(1)$ isometry (compatible with complex structure) by
assuming that the potential $K$ has one Killing symmetry, $R=Z+\bar
Z$:
\be
K=K(Z+\bar Z,Y_i,\bar Y_i,V_p,\bar V_p)\label{killing}
\ee
where $Z$ and $Y_i$ are chiral fields, whereas $V_p$ are twisted
chiral
fields. (Of course the discussion holds in the same way if
$Z$ is a twisted chiral field.)
In \cite{R,GHR} a duality transformation was described in which
twisted superfields are interchanged with untwisted ones.
Concretely, consider the `dual' potential
\be
\tilde K(R,Y_i,\bar Y_i,V_p,\bar V_p,\Psi+\bar\Psi)=
K(Z+\bar Z,Y_i,\bar Y_i,V_p,\bar
V_p)-R(\Psi+\bar\Psi),\label{dualpot}
\ee
where $Z$ is a chiral field and $\Psi$ a twisted chiral field.
Varying the action with respect to $\Psi$ gives back the original
theory. On the other hand one can equally well consider
the constraint coming from the variation with respect to $Z$,
\cite{R}
\be
{\delta S\over\delta Z}=0\quad\rightarrow\quad
{\partial K\over\partial r}-(\psi+\bar\psi)=0,\label{constraint}
\ee
and
the dual theory is obtained as a Legendre
transform of $K$. Now the independent variable are $\psi$, $y_i$
and $v_p$.
It follows that the dual metric has the following form:
\be
\tilde G_{\mu\nu}=\pmatrix{
0&-\tilde K_{\psi\bar\psi}&0&0&0&-\tilde K_{\psi\bar v_q}\cr
-\tilde K_{\psi\bar\psi}&0&0&0&-\tilde K_{v_p\bar\psi}&0\cr 0&
0&0&\tilde K_{y_i\bar y_j}&0&0\cr 0&0& \tilde
K_{y_i\bar y_j}&0&0&0
\cr 0&-\tilde K_{v_p\bar\psi}&
0&0&0&-\tilde K_{v_p\bar v_q}\cr -\tilde K_{\psi\bar v_q}&0&
0&0&-\tilde K_{v_p\bar v_q}&0\cr}.
\label{dualmetric}
\ee
Similarly, the dual antisymmetric tensor field is obtained as
\be
B_{\mu\nu}=\pmatrix{0&0&0&\tilde K_{\psi\bar y_i}&0&0\cr
0&0&\tilde K_{y_i\bar\psi}&0&0&0\cr
0&-\tilde K_{y_i\bar\psi}&
0&0&0&\tilde K_{y_i\bar v_p}\cr -\tilde K_{\psi\bar y_i}&0&
0&0&\tilde K_{v_p\bar y_i}&0
\cr 0&0&0&-\tilde K_{v_p\bar y_i}&0&0\cr 0&0&
-\tilde K_{y_i\bar v_p}&0&0&0\cr}.
\label{dualbmn}
\ee
Moreover, the dual dilaton field has the form
\be
2\tilde \Phi=2\Phi-\log 2K_{rr}.\label{dualphi}
\ee
It can be shown \cite{R,kkl}
that this N=2 duality transformation via
Legendre
transform is the same as the usual abelian duality transformation
\cite{B}.

\section{K\"ahler Spaces without Torsion and their Duals}

\setcounter{equation}{0}

It is already well known that if the torsion vanishes and there is no
dilaton field the condition that a $\s$-model has N=2 supersymmetry
is that the target space is K\"ahler, \cite{kahler}.
Thus, for the time being, we start with a K\"ahler manifold specified
(locally)
by its K\"ahler potential $K(u_{i},\ub_{i})$ and a dilaton field
$\P$.
The metric is given in terms of the K\"ahler potential by the
standard formula
\be
G_{ij}=G_{{\bar i}{\bar j}}=0\;\;\;,\;\;\; G_{i{\bar
j}}=K_{u_{i}\ub_{j}}
\label{kmetric}
\ee
It is obvious that the metric is invariant under the so called
K\"ahler transformations of the potential
\be
K(u_{i},\ub_{i})\rightarrow K(u_{i},\ub_{i})+\Lambda(u_{i})+{\bar
\Lambda}(\ub_{i})\label{ktran}
\ee
Then the Ricci-tensor takes its well-known form
\be
R_{u_i\bar u_j}=-\partial_{u_i}\partial_{\bar u_j}
U\;\;,\;\; R_{u_iu_j}=
R_{\bar u_i\bar u_j}=0\label{ric}
\ee
 with $U=\log\det K_{u_i\bar u_j}={1\over 2}\log\det G$.

The only condition for conformal invariance is $\beta_{\mu\nu}^G=0$
which here implies
\be
\Phi={1\over 2}U+f(u_i)+\bar f(\bar u_i),\label{kahl}
\ee
and
\be
\nabla_{u_i}\partial_{u_j}\Phi=\nabla_{\bar u_i}\partial_{\bar u_j}
\Phi=0.\label{phikah}
\ee
where $f$ is an arbitrary holomorphic function.
In addition we get
\be
\delta c={3\over 2}\alpha' K^{u_i\bar u_j}(8\partial_{u_i}\Phi
\partial_{\bar u_j}\Phi-4\partial_{u_i}\partial_{\bar
u_j}\Phi ).\label{dckah}
\ee

If one demands for enlarged $N=4$ world-sheet
supersymmetry, this implies that the K\"ahlerian space has to be
hyper-K\"ahler.
If the theory is also positive, then $\delta c=0$ from CFT arguments.
In such a case the Riemann tensor is self-dual and therefore the
space is Ricci-flat.
Ricci flatness and $\delta c=0$ implies constant dilaton.
However, we note that the hyper-K\"ahler condition is not
the only way to obtain $\delta c=0$; in fact we will provide
$N=4$ examples which are non-Ricci-flat and have non-constant
dilaton field. These examples will be presented later.

As described in \cite{kkl} it is not difficult to show that the
vanishing of the
 holomorphic double derivative on the dilaton implies,
 for non-trivial dilaton, that
there is
a generic Killing symmetry in the K\"ahler metric as well as in the
dilaton. Then, in a special coordinate system
the compatibility of the equations (\ref{kahl}) and
(\ref{phikah})
along with our
freedom to perform K\"ahler transformations implies that
\be
K=K(z+\bar z,y_{i},\bar y_{i})\;\;,\;\;\P=\d_{z}K=\d_{\bar z}
K\label{fe}
\ee
and
\be
\P={1\over 2}U+C(z+\bar z)\label{ff}
\ee
where $C$ is any real number.
We can take (\ref{fe}) as the equation specifying the dilaton in
terms of the
metric and then (\ref{ff}) becomes a non-linear differential equation
for the
K\"ahler potential
\be\det [K_{u_{i}\ub_{j}}]=\exp
[-2C(z+\bar z)+K_{z}+K_{\bar z}]
\label{fg}
\ee
generalizing the CY condition.

In the same coordinate system we can also compute
the
central charge deficit:
\be
\delta c=12\alpha' C.\label{centralcq}
\ee

Let us consider a special class
of solutions which can be regarded as the
generalization of the two-dimensional black hole backgrounds
found in \cite{bh}. Specifically, assume that the model
has a $U(N)$ isometry, i.e.
\be
K=K(x),\quad \Phi=\Phi(x) \qquad
x=\sum_{i=1}^N|u_i|^2.\label{unansatz}
\ee
The general form of the metric is then
\be
K_{u_i\bar u_j}=K'\delta_{ij}
+K''\bar u_i u_j,\qquad K'={\partial K\over\partial
x}.\label{unisom}
\ee
For $N>1$, the linear term in the dilaton, eq. (\ref{ff}),
is not allowed by the $U(N)$ isometry and the dilaton field becomes
\be
\P={1\over 2}U={1\over 2}\log
\lbrack(K')^{N-1}(K'+xK'')\rbrack.\label{dilisom}
\ee
Let us define the following function:
\be
Y(x)=xK'(x).\label{yfct}
\ee
Now we have to insert the ansatz eqs. (\ref{unisom}, \ref{dilisom})
into the field equation (\ref{phikah}), and the solution of this
equation takes the following form:
\be
e^{Y}\sum_{m=0}^{N-1}{(-1)^mY^m\over m!}=A+Bx^N.\label{solution}
\ee
Here $A$ and $B$ are arbitrary parameters.
{}From this we immediately obtain
\be
Y'=BN!(-1)^{N-1}e^{-Y}Y^{1-N}x^{N-1}.\label{yprime}
\ee
Then the dilaton, eq. (\ref{dilisom}), can be also expressed
entirely of $Y$ as
\be
\P={1\over 2}U=-{1\over 2}Y+{\rm const.}\label{dily}
\ee
The Ricci tensor becomes
\be
R_{u_i\bar u_j}=-
\partial_{u_i}\partial_{\bar u_j}U=
Y'\delta{ij}+Y''\bar u_i u_j.\label{ricciy}
\ee
The scalar curvature can be computed to be
\be
R=2(N-xY')=2(N-f(Y))
.\label{scalarcy}
\ee
with
\be
f(Y)=Y^{1-N}\biggl\lbrack
\sum_{m=0}^{N-1}{(-1)^mY^m\over m!}-Ae^{-Y}\biggr\rbrack (-1)^{N-1}N!
\ee
Finally, the central can be expressed as
\be
\delta c=3N\alpha'.\label{centralun}
\ee

The explicit form of the scalar curvature, eq. (\ref{scalarcy}),
allows
us to discuss the asymptotic behavior and the singularity
structure of our class of solutions. First we recognize
that the $2N$-dimensional K\"ahler space
has zero scalar curvature for $Y\rightarrow \infty$.
Second, for all $N$, the scalar curvature possesses a generic
singularity for $Y\rightarrow -\infty$.
Moreover, there is another singularity at $Y=0$ for $N>1$
if $A\neq 1$. On the other hand, if $A=1$, the scalar curvature is
regular at $Y=0$ and becomes $R(Y=0,A=1)=2N$.

For the simplest case, namely two-dimensional backgrounds with $N=1$,
the solution (\ref{solution}) reproduces the well-known backrounds
  which correspond to the exact conformal field theories
$SL(2,R)/U(1)$ and $SU(2)/U(1)$ respectively depending on the
choice of the parameters $A$ and $B$. Moreover the duality
transformation
(\ref{dualpot}) \cite{G} exactly corresponds to axial versus vector
gauging of
$U(1)$ in the corresponding conformal field theory \cite{K,DVV}.

Let us now consider four-dimensional backgrounds which are not direct
products of two-dimensional spaces. Specifically we consider
solutions of the form eq. (\ref{solution}) with
$N=2$. One has to emphasize that so far it is not known to us
which exact supercoformal field theory might correspond to
this type of backgrounds.
Using $Y$ together withe the overall phase $\theta$ as (real)
coordinates,
the metric then reads:
\be
\eqalign{{\rm d}s^2&={({\rm d}Y)^2\over  4f(Y)}+{f(Y)\over 4}
\biggl({\rm d}\theta
-i{\bar y{\rm d} y-y{\rm d}\bar y\over 1+y\bar y}\biggr)^2\cr &
+{Y\over (1+
y\bar y)^2}
{\rm d}y{\rm d}\bar y,\quad
f(Y)={2(Ae^{-Y}+Y-1)\over Y}.\label{betterme}\cr}
\ee
This metric  in the ($\theta,\psi,\phi$)
subspaces is a deformation of the fibration of $S^{3}$ over
$S^2$, whose line element is manifest in (\ref{betterme}).

Now we will analyse the structure of the Euclidean manifold as a
function
of A and B.
We need some asympotics of the function $f(Y)$:
\be
f(\infty)=2 \;\;,\;\;f(-\infty)=-{\rm Sign}[A]\times \infty\label{f1}
\ee
\be
f(0^{+})=-f(0^{-})={\rm Sign}[A-1]\times \infty\label{f2}
\ee
Since $x^2$ must be positive, we are dealing with the following
cases:

1) $A>1,B<0$. There are two manifolds, the first with $Y\geq 0,
x^2>(1-A)/B$
with signature (4,0)
and a curvature singularity at $Y=0$ and the second with$Y\leq 0,
(1-A)/B\leq x^2 \leq -A/B$ with signature (0,4)
and curvature singularities at $Y=0,-\infty$.

2) $A=1, B<0$. There is a regular Euclidean (4,0) manifold for $Y>0,
x^2 >0$
and a
singular (at $Y=-\infty$) Eclidean (0,4) manifold for $Y<0, 0\leq x^2
\leq
-1/B$.

3) $0<A<1$. In this case $f(Y)$ has a positive and a negative zero
which
we will denote by $Y_{\pm}$: $f(Y_{\pm})=0$.
For $B<0$ there is again a regular (finite curvature) Euclidean
manifold for
$Y>Y_{+}$ with signature (4,0) and another with $Y<Y_{-}$ with
signature (0,4)
and a curvature singularity at $Y=-\infty$.
For $Y_{-}<Y<Y_{+}$ and $B>0$ there is another singular  manifold
with
signature (2,2).

4) $A\leq 0$. In this case $f$ has a single positive zero, $Y_{+}$.
For $B<0$ and $Y>Y_{+}$ we have a regular manifold with signature
(4,0).
For $B>0$ and $Y<Y_{+}$ there is a singular manifold with signature
(2,2).

Applying eq. (\ref{dualpot}), the dual metric is given
 as
\be
{\rm d}\tilde s^2={
{\rm d}\psi{\rm d}\bar\psi\over f(Y)}
+{Y\over (1+y\bar y)^2}{\rm d}y{\rm d}
\bar y,
\label{dualntwo}
\ee
whereas the dual dilaton and antisymmetric tensor field are as
follows
$
\tilde\Phi=-{1\over 2}\log[e^Yf(Y)]$, $\tilde B_{\psi
\bar y}=2y/(1+y\bar y)$.
The dual scalar curvature becomes
\be
\tilde R ={2Y^2(ff''-f'^{2})+f^{2}+4Yf\over Y^2 f}\label{dualsc}
\ee

Thus, for  the dual space, $\tilde R =0$  for $Y\rightarrow\infty$,
and there are curvature singularities at $Y=-\infty,0$ and, for
generic values
of $A$ at the zeros of $f(Y)$ (e.g. for $A=-1$,
at $Y\simeq 1.3$).

\section{Four-dimensional Non-K\"ahlerian Spaces with Torsion and
their Duals}

\setcounter{equation}{0}

The discussion in the case of non-vanising antisymmetric tensor
fields will be restricted to the simplest non-trivial case, namely
four-dimensional
target spaces, i.e. $m=n=1$.
In that case it can be shown that the solutions fall into three
mutually exclusive classes \cite{kkl}:

i) Solutions whose quasi-K\"ahler potential satisfies the ordinary
Laplace equation,
\be
(\partial_u\partial_{\bar u}+\partial_v\partial_{\bar
v})K=0.\label{laplace}
\ee
and the dilaton field is simply given as
\be
2\Phi=\log K_{u\bar u}+{\rm constant}.\label{dilnfour}
\ee

ii) Solutions with one isometry whose quasi-K\"ahler potential
satisfies
\be
K_{ww}=K_{u\bar u}e^{-K_w+c_1(w+\bar w)+c_2}.\label{solveb}
\ee
in a special coordinate system
and
\be
2\Phi=\log(K_{v\bar v})-c_1(w+\bar w)-c_2\label{solvea}
\ee

ii) Solutions with two isometries whose quasi-K\"ahler potential
satisfies
\be
K_{ww}e^{K_w+c_{2}(w+\bar w)}=K_{zz}e^{K_z+c_1(z+\bar
z)}.\label{finalcon}
\ee
in a special coordinate system, and
\be
2\Phi=\log K_{zz}-K_w-c_1(z+\bar z)+{\rm constant}.\label{dilfin}
\ee
In the above $c_{1,2}$ are constants.

In the following we will focus on the solutions of case (i).
Eqs.(\ref{laplace}) and (\ref{dilnfour})
imply that $\delta c=0$ and these backgrounds are expected to have
$N=4$
superconformal symmetry.  This obeservation
is consistent with the fact that eq.(\ref{laplace})
is the generalization of the hyper-K\"ahler
condition for spaces with antisymmetric
tensor field. The form of the dilaton field
has the important consequence that the four-dimensional
metric in the Einstein frame is flat:
$
G_{\mu\nu}^{{\rm Einstein}}=e^{-2\Phi}G_{\mu\nu}^\sigma=\delta_{\mu
\nu}$.
In fact,
the solutions
of the dilaton equation (\ref{dilnfour}) are
the type II versions of the axionic solutions of \cite{rey}:
\be
{\rm d}\Phi=\pm {1\over 2}e^{-2\Phi}H^*.\label{selfdual}
\ee
This relation is nothing else than the self-duality
condition on the dilaton-axion field. Its solutions are
known as axionic instantons.
All these solutions leave spacetime supersymmetry unbroken.
In particular it can be shown that one of the solutions in this class
(which in its heterotic version was identified with a magnetic
monopole
background \cite{khuri}), turns out to be a dual of flat space
\cite{kkl}.

The form of the solutions of the Laplace equation depends on the
number of isometries of the theory (which are compatible with the
complex structure). In the case with two
translational $U(1)$ Killing symmetries, i.e. $K=K(u+\bar u,v+\bar
v)$
the most general solution of (\ref{laplace}) looks like
\be
K=iT(u+\bar u+i(v+\bar v))-i\bar T(u+\bar u-i(v+\bar v)).\label{f}
\ee
In the case with one traslational isometry the general solution
becomes
\be
K(u+\bar u,v,\bar v)=i\int{\rm d}\beta T(\beta,v+\beta(u+\bar u)
-\beta^2\bar v)+{\rm c.c.}\label{g}
\ee
where in both (\ref{f}), (\ref{g}) T is an otherwise arbitrary
function.

Let us now
construct the dual spaces for the solutions
of the Laplace equation with one or two isometries,
(\ref{f},\ref{g}).
We will perform a duality transformation  on the chiral $U$-field
replacing it by a twisted chiral field $\Psi$. The Legendre
transformed potential $\tilde K$ will only contain twisted
fields and will be therefore a true K\"ahler function
leading to a non-compact K\"ahler space without torsion.

Doing the Legendre transform we obtain the following line element
\be
{\rm d}s^2={1\over K_{uu}}({\rm d}z-K_{uv}{\rm d}v)({\rm d}{\bar
z}-K_{u\bar
v}{\rm d}{\bar v})-K_{v\bar v}{\rm d}v{\rm d}{\bar v}
\label{riflat}
\ee
where $K(u+\bar u ,v,\bar v)$ is the original quasi-K\"ahler
potential that satisfies the Laplace equation $K_{uu}+K_{v\bar v}=0$
and $z,\bar z$ are the dual coordinates defined via the Legendre
transform $z+\bar z =K_{u}$.
The coordinates $v,\bar v, z,\bar z$ are now the K\"ahler
coordinates.
The Laplace equation implies that the determinant of the K\"ahler
metric
(\ref{riflat}) is constant so we obtain a Ricci flat K\"ahler
manifold.
The dual dilaton is consequently constant.

The general solution to the 4-d Laplace equation with one isometry
can be
written as in (\ref{g}).
Let us introduce the notation
\be
<T>\equiv \int{\rm d}\beta T(\beta,v+\beta(u+\bar u)-\beta^2\bar v).
\label{nota}
\ee
and the function
\be
Z(u+\bar u,v,\bar v)=K_{u}=i<\beta(T_{v}-\bar T_{\bar v})>
\label{def}
\ee
and we should remember that $z+\bar z=Z(u+\bar u,v,\bar v)$.
Then the line element (\ref{riflat}) can be written in the form
\be
{\rm d}s^2={1\over G}({\rm d}z-A_{v}{\rm d}v)({\rm d}{\bar z}-\bar
A_{\bar
v}{\rm d}{\bar v})+G{\rm d}v{\rm d}{\bar v}
\label{bun}
\ee
where,
\be
G={\partial Z\over \partial u}\;\;,\;\;A_{v}={\partial Z\over
\partial
v}\;\;,\;\;{\bar A}_{\bar v}={\partial Z\over \partial \bar v}
\label{def1}
\ee
The interpretation of the metric (\ref{bun}) is as follows:
The $G{\rm d}v{\rm d}{\bar v}$ part describes the metric of a 2-d
Riemann
surface (generically non-compact). The metric depends also on $z+\bar
z$.
For fixed $z+\bar z$, $A_{v},{\bar A}_{\bar v}$ describe a flat line
bundle on the Riemann surface.
The metric (\ref{bun}) is that of a flat complex line bundle on the
Riemann
surface.
The functions $G$, $A_{v},{\bar A}_{\bar v}$ are harmonic.

The metric (\ref{bun}) describes a large
class of 4-d non-compact Calabi-Yau manifolds, which are also
hyper-K\"ahler.
The associated $\s$-models have N=4 superconformal symmetry and $c=6$
($\tilde
c =2$).
The manifolds have generically asymptotically flat regions as well
as curvature singularities.

Let us briefly display a simple example choosing
\be
T=-i\gamma(\beta)e^{u+\bar u+{v\over\beta}-\beta\bar v}.\label{exg}
\ee
Then the potential becomes
\be
K=e^{u+\bar u}\phi(v,\bar v),\qquad \phi(v,\bar v)=
\int{\rm d}\beta\gamma(\beta)\lbrack e^{{v\over\beta}-\beta\bar v}+
e^{{\bar v\over\beta}-\beta v}\rbrack .\label{exga}
\ee
In turn, the dual space is determined by the following
K\"ahler potential:
\be
\tilde K=(z+\bar z)\log(\psi+\bar\psi)-(\psi+\bar\psi)\log
\phi(v,\bar v).\label{dualexg}
\ee
The intergral in (\ref{exga}) can be explicitly performed if we
choose
$\gamma(\beta)=e^{-A\over\beta}\beta^{\nu-1}$:
\be
\phi(v,\bar v)={\rm constant} \biggl(\sqrt{{A-v\over\bar
v}}\biggr)^\nu
K_\nu(2\sqrt{(A-v)\bar v})+{\rm h.c.}\label{integral}
\ee
Here $K_\nu$ is the Bessel function with complex argument.

Let us study now the (more symmetric) special case of (\ref{bun})
with two isometries, i.e. $K(u+\bar u,v+\bar v)$.
If we paramertrize, $u=r_{1}+i\theta$, $v=r_{2}+i\phi$ then K is of
the form
$K(r_{1},r_{2})=iT(r_{1}+ir_{2})-i\bar T(r_{1}-ir_{2})$.
Introducing a new complex coordinate $z=r_{1}+ir_{2}$, we can write
the metric
(\ref{bun}) in the following suggestive form
\be
{\rm d}s^{2}={Im{\rm T}\over 2}{\rm d}z{\rm d}{\bar z}+{2\over Im{\rm
T}}
({\rm d}\theta +{\rm T}{\rm d}\phi)({\rm d}\theta +{\bar {\rm T}}{\rm
d}\phi)\label{tor}
\ee
where T$(z)$ is an arbitrary meromorphic function.
It is crucial to note that the metric (\ref{tor})
is $not$ written in K\"ahler coordinates.
Such coordinates are $v,\bar v$ and $w,\bar w$ with $w+\bar
w=iT'(r_{1}+ir_{2})-i\bar T'(r_{1}-ir_{2})$ and $w-\bar w =2i\theta$.

Now the interpretation of the metric (\ref{tor}) is straightforward:
If we take $\theta,\phi$ to be angular variables, then they
parametrize
a 2-d torus, with modulus T$(z)$ which depends holomorphically
on the rest of the coordinates and conformal factor proportional
to $1/Im $T.
The zeros and poles of the Riemann tensor are determined by the zeros
and
poles (or essential singularities) of the function T$(z)$.

This solution (with a different interpretation) was found in
\cite{NCCY},
where some global issues were also addressed\footnote
{Some generalizations of this idea to more dimensions were recently
presented
in \cite{hu}.}.
We should note that as in \cite{NCCY} a full invariance under the
torus
modular group, $T\rightarrow T+1$ and $T\rightarrow -1/T$ can be
implemented
by a holomorphic coordinate transformation in $z$, which will modify
$Im$T to a modular invariant in the first part of (\ref{tor}).
It was also argued that such a metric might receive higher order
corrections.
However we have just shown that this metric is the dual of the family
of
wormhole solutions which are absolutely stable as CFTs due to their
N=4
superconformal symmetry and it does possess a hyper-K\"ahler
structure
although not easily visible in this coordinate system.

The 4-d non-compact CY manifolds presented in this section constitute
a
large class of exact solutions to superstring theory with extended
supersymmetry.
A detailed analysis of their structure as well as their potential
Minkowski
continuations is beyond the scope of this work and is reserved for
future
study.

\section{Conclusions}
\setcounter{equation}{0}

We have examined some four-dimensional superconformal theories with
N=2 and N=4
superconformal symmetry (classical solutions to superstring theory).
We show that there exists a plethora of such theories with
non-trivial
metric, dilaton and antisymmetric tensor field.

Our solutions are classified in two classes:
(i) Those that are based on a K\"ahler manifold (when
$H_{\mu\nu\rho}=0$).
(ii) Non-K\"ahlerian solutions with non-zero torsion.
These two subclasses are related by $Z_{2}$ duality transformations
(when
isometries are present).
$Z_{2}$ duality interchanges the roles of untwisted and twisted
chiral
superfields and act in a manifest N=2 preserving fashion.

In the K\"ahlerian case we show that the presence of a non-trivial
dilaton field implies the presence of an isometry in the background
data
(K\"ahler metric and dilaton).
Among the K\"ahlerian solutions we find a large class of
(non-compact)
Ricci-flat (CY) manifolds with one isometry.
This class of solutions generalizes the compact 4-d Ricci flat
manifolds (K3).
A special case of the solutions above (with two isometries) is that
of ref.
\cite{NCCY} found in a slightly different context.
These CY manifolds are duals of non-zero torsion solutions with N=4
superconformal symmetry.

Let us finally emphasize that it is
a very interseting problem to find the exact $N=2$ and $N=4$
superconformal field theories which correspond to our general
solutions.
Upon analytic continuation of the Euclidean solutions
we expect to obtain many cosmological solutions to superstring theory
whose
spacetime
properties deserve further study.

\vskip .5cm
\noindent
{\bf Acknowledgments} \\
This work was partially supported by EEC grants, SC1$^{*}$-03914C
and SC1$^*$-CT92-0789.
\noindent

\def\PL{Phys. Lett. }
\def\NP{Nucl. Phys. }
\def\PR{Phys. Rev. }

\end{document}